# AMSDU vs AMPDU: A Brief Tutorial on WiFi Aggregation Support

Gautam Bhanage[1]
Tech Report*: Bhanage.com / GDB2017-004
gbhanage@gmail.com
April 2017

*Abstract*— **WiFi MAC architecture supports aggregation at two layers. The MAC service data units (MSDUs) can be aggregated to form AMSDUs. Each AMSDU serves as a single MAC protocol data unit (MPDU). Another layer of aggregation is introduced when MPDUs can be packed together to deliver an aggregated MPDU (AMPDU). In this brief tutorial we discuss (1) why aggregation is useful at all, (2) why two layers of aggregation are required, (3) trade-offs for aggregating at different layers.**

*Index Terms*— **WiFi, WLAN systems, Aggregation, MAC mechanisms, Optimization, Tutorial.**

## I. Introduction

**W**hat *is aggregation and why bother aggregating*? Well as the term suggests, aggregation, combines multiple packets which would have been transmitted independently into a single transmission unit. Why should wireless drivers bother to aggregate packets? As shown in previous studies [2], the wireless medium has an inherently high overhead associated with every physical layer protocol data unit (PPDU), commonly referred to as a single wireless transmission. Every transmission incurs the cost of preamble, physical layer headers, MAC headers and inter-frame spaces (IFSs). These overheads are constant and are incurred for every physical transmission[2]. Thus if we reduce the number of individual transmissions by combining frames together, we will incur lesser over heads in transmission. As shown in Figure 1, the results from [2] indicate there are significant savings even with the aggregation of just 2 frames together at rates as low as 54Mbps. In the Figure 1, to understand the benefits, please compare the fraction of channel utilization available for the payload as compared to the overheads. It has been shown that with higher physical layer transmission rates, these savings are even more pronounced.

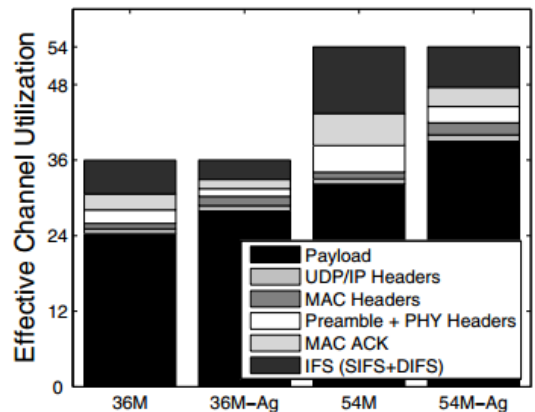

*Figure 1: Savings in airtime used on the channel by using aggregation as shown in [2].*

## II. To aggregate or not to aggregate

So we all agree that aggregation improves transmission efficiency. Then **why not aggregate everything every time?** What is the downside of aggregating? Mainly, two issues: (1) Delay – if the transmitter waits to collect frames to aggregate, and

---

(2) Higher packet error rate (PER) – this is potentially the case if the transmitter aggregates frames increasing the packet size of a single transmission.

To counter delays, most WiFi transmitters will institute a max-timeout to limit delays. These timeouts might be different for different traffic types. *The PER containment is done by supporting 2 layers of aggregation.*

### III. WHY HAVE TWO TYPES OF AGGREGATION

As shown in the Figure 2 below, WiFi protocol supports aggregation at 2 layers. Multiple MSDU packets can be combined into an AMSDU. This AMSDU unit serves as one packet as passed down by higher layers to the MAC. The CRC is calculated on each of these AMSDUs. So if any single AMSDU transmission fails, the entire AMSDU has to be re-transmitted. *Thus the effective PER for a considered BER is determined by the size of the AMSDU.*

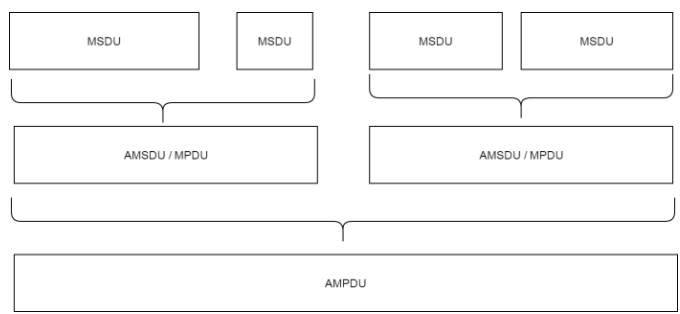

*Figure 2:WiFi Two Layer Aggregation*

However, **if the protocol supported only AMSDU layer of aggregation, the benefits of aggregation achieved by aggregating multiple MAC layer units would have been limited by the achievable PER for the aggregate size.** Instead, the WiFi protocol allows the sender to aggregate multiple AMSDU (also referred to as MPDUs) units into a single AMPDU while allowing CRC checks and retries for each AMSDU within an AMPDU. Thus the WiFi protocol allows us to achieve higher MAC efficiency by transmitting AMPDUs while limiting PERs and re-transmissions at the AMSDU level.

So, **why not do just AMPDUs and skip AMSDUs?** This is feasible. However, including AMSDUs as a part of AMPDUs is more efficient because this results in: (1) Fewer CRC calculations for smaller packet sizes at sender and receiver – once per AMSDU as opposed to once every MSDU and (2) Fewer MAC headers (MSDU headers). These benefits might seem small. However, when we consider that most of the internet traffic consists of TCP transmissions with small packet sizes [3], these small savings can add up considerably.

Another advantage of doing AMSDUs is as shown in Figure 3. While doing AMSDU aggregation, in the 802.11 QOS header, the source and transmitter address are the radio's transmitter address. Similarly the destination address is made to match the receiver address. The reason this is done because, **a single AMSDU can contain MSDUs with potentially different source and destination addresses as long as they are of the same traffic type (TID).**

```
▲ IEEE 802.11 QoS Data, Flags: ......F.
    Type/Subtype: QoS Data (0x0028)
  ▷ Frame Control Field: 0x8802
    .000 0000 0011 0000 = Duration: 48 microseconds
    Receiver address:                    46:5c)
    Destination address:              5a:46:5c)
    Transmitter address:              0c:9c:00)
    Source address:                      9c:00)
    BSS Id:                           :9c:00)
    STA address:                      5a:46:5c)
    .... .... .... 0000 = Fragment number: 0
    1100 0010 0011 .... = Sequence number: 3107
  ▷ Qos Control: 0x0080
▲ IEEE 802.11 Aggregate MSDU
  ▲ A-MSDU Subframe #1
      Destination address:            5a:46:5c)
      Source address:                 71:77:62)
      A-MSDU Length: 1506
    ▷ Logical-Link Control
    ▷ Internet Protocol Version 4, Src: .         l.com (172.
    ▷ User Datagram Protocol, Src Port: 64063 (64063), Dst Port: 
    ▷ Data (1470 bytes)
  ▷ A-MSDU Subframe #2
  ▷ A-MSDU Subframe #3
```

*Figure 3: AMSDU Aggregation as seen on a wireless sniffer capture*

### IV. AMSDU VS AMPDU CRITERIA

The size limit on the AMSDU and the AMPDUs are based on negotiations between the transmitter and the receiver. The first level of negotiation happens in the assoc-request and assoc-response frames. Here the support for AMSDU is negotiated with the maximum frame size. Each entity advertises its capability as 3.8K, 7.9K or 11K byte size AMSDU support. The aggregation support is hashed out in the add-block-ack request and add-block-ack response frames. These negotiate the size of the window for the AMPDU transmissions for a certain TID.

Once the maximum sizes are negotiated, the individual transmission sizes are dependent on implementation. Multiple studies have proposed different mechanisms for dual layer aggregation [1][4].

**AMSDU aggregation criteria:** Some factors which come to mind are (1) frame size – benefits are more for smaller frame sizes, (2) timeout – to limit delays seen on the air due to aggregation, (3) traffic type – we might want to limit the amount of aggregation and hence delay – jitter seen with certain traffic types, (4) Achievable PER – there is an optimal point at which the cost of retransmissions will be much less as compared to the benefit achieved by AMSDU aggregation.

**AMPDU aggregation criteria:** For AMPDUs, the same set of factors play a role in the decision making process as those seen for AMSDU aggregation. However, PER is not a consideration here since each individual aggregated unit can now be independently retried.

Some fancy aggregation scheduling mechanisms should be able to jointly determine the optimal solution for aggregating at both layers. In other cases, due to the time complexity of computing the solution at run time on a very granular basis, some or most real time schedulers use heuristics to make these decisions. For example, most systems might make the AMSDU size decision statically. Based on capacity negotiated with the receiver i.e. the size of the aggregate and the number of units in each aggregate, the sender might use a simple timeout for independently making AMSDUs and feed these to the AMPDU aggregation unit. Such algorithms will rely on the rate control and transmit power control mechanism to manage the PER for the link[3].

V. CONCLUSION & FUTURE DIRECTIONS

This brief tutorial provides concise justification behind the use of a two layer aggregation scheme used in WLANs. It provides a brief insight into the "why" behind some aggregation mechanisms used in a WiFi transmitter.

---

[3] Some might argue that dropping rates has an adverse effect on the link performance as compared to just losing some aggregating efficiency. Justifying this decision is not within the scope of this paper